\documentclass[]{interact}

\usepackage{epstopdf}
\usepackage[caption=false]{subfig}

\usepackage[numbers,sort&compress]{natbib}
\bibpunct[, ]{[}{]}{,}{n}{,}{,}
\makeatletter
\def\NAT@def@citea{\def\@citea{\NAT@separator}}
\makeatother

\theoremstyle{plain}

\theoremstyle{definition}

\theoremstyle{remark}

\usepackage{listings, jvlisting}

\renewcommand{\theenumi}{\roman{enumi}}

\begin{document}

\title{NIMS-OS: An automation software to implement a closed loop between artificial intelligence and robotic experiments in materials science}

\author{
\name{Ryo Tamura\textsuperscript{a,b}\thanks{CONTACT Ryo Tamura. Email: tamura.ryo@nims.go.jp},
Koji Tsuda\textsuperscript{b,a}\thanks{CONTACT Koji Tsuda. Email: tsuda@k.u-tokyo.ac.jp},
and Shoichi Matsuda\textsuperscript{c,d}\thanks{CONTACT Shoichi Matsuda. Email: matsuda.shoichi@nims.go.jp}
}
\affil{
\textsuperscript{a}Center for Basic Research on Materials, National Institute for Materials Science, Tsukuba, Japan; \\
\textsuperscript{b}Graduate School of Frontier Sciences, The University of Tokyo, Chiba, Japan; \\
\textsuperscript{c}Research Center for Energy and Environmental Materials(GREEN), National Institute for Materials Science, Tsukuba, Japan; \\
\textsuperscript{d}Center for Advanced Battery Collaboration, Research Center for Energy and Environmental Materials (GREEN), National Institute for Materials Science, Tsukuba, Japan
}
}

\maketitle

\begin{abstract}
NIMS-OS (NIMS Orchestration System) is a Python library created to realize a closed loop of robotic experiments and artificial intelligence (AI) without human intervention for automated materials exploration. 
It uses various combinations of modules to operate autonomously.
Each module acts as an AI for materials exploration or a controller for a robotic experiments.
As AI techniques, Bayesian optimization (PHYSBO), boundless objective-free exploration (BLOX), phase diagram construction (PDC), and random exploration (RE) methods can be used.
Moreover, a system called NIMS automated robotic electrochemical experiments (NAREE) is available as a set of robotic experimental equipment. 
Visualization tools for the results are also included, which allows users to check the optimization results in real time. 
Newly created modules for AI and robotic experiments can be added easily to extend the functionality of the system. 
In addition, we developed a GUI application to control NIMS-OS.
To demonstrate the operation of NIMS-OS, we consider an automated exploration for new electrolytes.
NIMS-OS is available at \texttt{https://github.com/nimsos-dev/nimsos}.
\end{abstract}


\section{Introduction}

The integration of artificial intelligence (AI) and robotic experiments is essential to realize automated materials exploration. 
If an AI system can take on some information tasks conventionally performed by human researchers and robotic systems can then execute the required physical tasks,
experiments for materials exploration can proceed automatically.
Such a platform may be expected to discover many novel materials and lead to substantial innovation in materials science.
In recent years, significant progress has been made in the development of AI techniques and robotic devices suitable for materials exploration.

Since the launch of the Materials Genome Initiative\cite{White:2012uz}, machine learning and AI have been actively used for materials exploration\cite{Ramprasad:2017aa,Schmidt:2019aa,Ling:2022aa}.
In general, materials exploration can be regarded as the problem of finding optimal materials from among a materials search space.
The elements to be used in the search space must be configured, along with its composition range, process parameter range, and so forth.
To solve this problem, 
black-box optimization methods are useful\cite{terayama_2021}, 
and various methods have been developed and applied to fit various needs.
Bayesian optimization (BO) is among the most frequently used methods in materials science\cite{UENO201618}.
In this method, promising materials can be selected in the materials search space using the predictions of their properties and the uncertainty of these predictions evaluated by Gaussian process regression.
Using BO,
various real materials, such as Li-ion conductive materials\cite{Homma-2020}, multilayered metamaterials\cite{doi:10.1021/acscentsci.8b00802}, halide perovskite\cite{SUN20211305}, superalloys\cite{Tamura-2021}, and electrolytes\cite{MATSUDA2022100832}, have been explored. 
BO is also used for the automation techniques for analysis of materials\cite{Ozaki:2020aa,doi:10.1080/27660400.2022.2146470}.
In addition, many methods have been proposed for black-box optimization in materials exploration, such as Monte Carlo tree search\cite{doi:10.1080/14686996.2017.1344083}, rare event sampling\cite{D1DD00043H}, and algorithms using an Ising machine\cite{kitai2020designing,doi:10.7566/JPSJ.90.064001,PhysRevResearch.4.023062}. 
In the future, many more innovative methods are expected to be developed.

Robotic experiments have progressed to realize laboratory automation of chemical analysis and high-throughput screening in the field of biology\cite{doi:10.1126/science.1165620,doi:10.1126/science.295.5554.517,Olsen:2012aa,Macarron:2011aa}.
Various types of automated analyzers and pipetting devices have been developed, 
and robotic arms have been used as a transport system to connect these systems.
Moreover, robotic technology has been used to explore novel materials, such as thin-film materials\cite{doi:10.1126/sciadv.aaz8867,doi:10.1063/5.0020370}, battery electrolytes\cite{DAVE2020100264,MATSUDA2022100832}, and photocatalysts\cite{Burger:2020uf}. 
These studies used BO to automate the proposal of promising experimental conditions.
This enables a closed loop of robotic experiments and AI that can perform automated materials exploration without human intervention.
This approach involves some key advantages, such as the ability to generate materials data of uniform quality and the absence of human error. 
In contrast, 
at present,
robotics systems are limited in their ability to perform complex material synthesis tasks that require the skills of experts.
Thus, further innovation in robotic devices will be important.

In addition to AI and robotic technologies, 
the control systems and software used to interlink them are also an important element to realize a closed loop without human intervention. 
Generally, different AI algorithms should be used depending on the motivation of a materials exploration task. 
Furthermore,
the procedure to control the devices should depends on the nature and characteristics of the robotic systems used. 
Therefore, control software has thus far been developed on a case-by-case basis for different AI algorithms and robotic systems.

\begin{figure}
\centering
\includegraphics[scale=0.5]{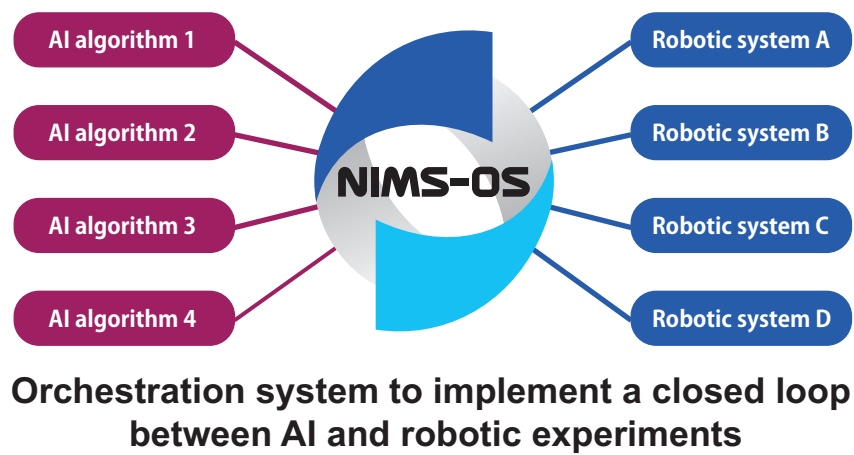}
\caption{
Image of the combinations of AI algorithms and robotic systems via NIMS-OS.
}
\label{fig:modules}
\end{figure}

In this study,
we developed NIMS-OS (NIMS Orchestration System) to realize a closed loop between AI models and robotic experiments, 
with the aim of establishing a generic control software system.
Although this software was written in the Python programming language, 
we also developed a GUI version to improve operability after installation.
NIMS-OS treats each AI algorithm and each robotic system as separate modules (see Fig.~\ref{fig:modules}). 
This enables the implementation of a closed loop with any combination of these modules.
If modules for new AI algorithms or robotic systems are prepared,
new closed-loop systems can be easily controlled via NIMS-OS.
One of the advantages of developing such generic control software is the establishment of technical standards for automated materials exploration. 
For AI algorithms, we determined standard formats for the input and output.
Algorithms created according to the standard format can be immediately tested using any currently available robotic system. 
We expect this work to contribute to the development of new AI algorithms for automated materials exploration.
For robotic systems,
we expect modules developed based on NIMS-OS to increase the commonality of operational procedures,
leading to cost reductions as new robotic experimental devices are introduced. 
Note that ChemOS\cite{10.1371/journal.pone.0229862} is similar to NIMS-OS; it was developed as an automation system in the field of chemistry.

\begin{figure}
\centering
\includegraphics[scale=0.6]{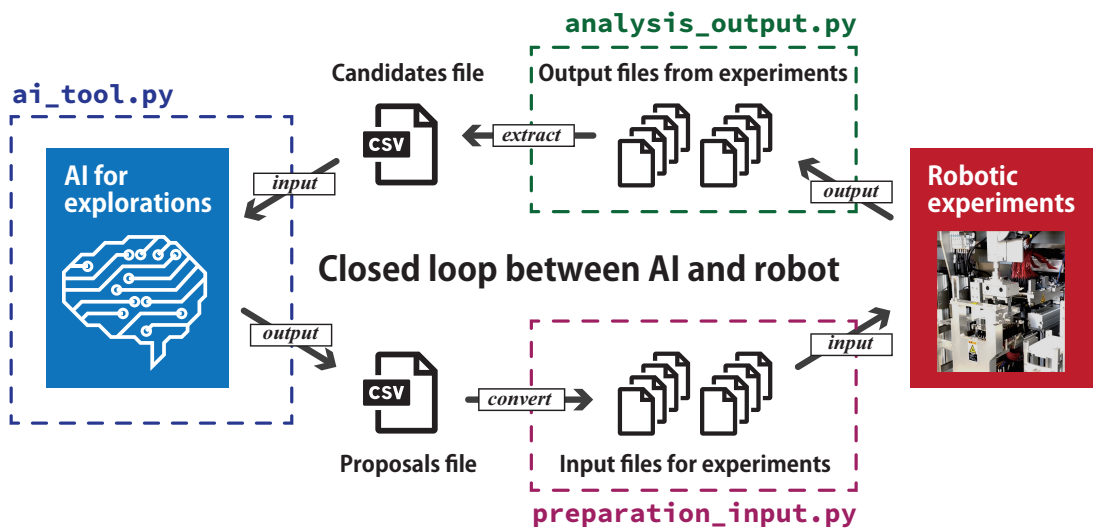}
\caption{
Procedures in NIMS-OS and roles of each Python scripts.
}
\label{fig:cycle}
\end{figure}

Let us briefly introduce the specifications of NIMS-OS.
First,
a candidates file listing experimental conditions as a materials search space should be prepared in advance.
A closed loop is formed according to the following three steps (see Fig.~\ref{fig:cycle}):
\begin{itemize}
\item[Step 1:] Select promising experimental conditions from the candidates file using an AI model.

\item[Step 2:] Create an input files for the robotic experiments and execute the experiments.

\item[Step 3:] Analyze the output from the experiments and update the candidates file based on the experimental results.
\end{itemize}
Currently, the following AI algorithms are used as modules, which are available for Step 1: (i) Bayesian optimization using PHYSBO\cite{MOTOYAMA2022108405}, (ii) boundless objective-free exploration using BLOX\cite{D0SC00982B}, and (iii) phase diagram construction using PDC\cite{PhysRevMaterials.3.033802} and random exploration.
Approach algorithms can be selected according to the purpose of materials exploration effort.
For Steps 2 and 3,
a reference module is provided for robotic experiments, which enables operation checks even without devices, along with a module for NIMS automated robotic electrochemical experiments (NAREE).
We plan to continue developing additional modules for this system.

The reminder of this study is organized as follows.
Section~\ref{sec:preparation} describes the preparation of a candidates file storing experimental conditions. 
In Section~\ref{sec:modules}, we introduce the available modules for the AI and robotic experiments in NIMS-OS.
Section \ref{sec:python} details the use of the Python code, and the usage of the GUI version is explained in Sec.~\ref{sec:gui}. 
In Section~\ref{sec:app}, as a demonstration, 
the results of an autonomous electrolyte exploration via a closed-loop approach using PHYSBO and NAREE in NIMS-OS are described.
Finally,  Section \ref{sec:summary} concludes this work with some discussion and suggest some important avenues for further research.

\section{Preparation of candidates for experimental conditions} \label{sec:preparation}

A major feature of NIMS-OS is that a data file listing candidate experimental conditions is prepared in advance (we refer to this data file a candidates file).
In general,
because there are many candidates,
conducting experiments in all possible conditions is impractical.
Thus, automated materials exploration proceeds by selecting promising experimental conditions from these listed candidates.
This makes the closed-loop strategy more generalizable.
That is, a variety of exploration motivations and robotic systems can be handled by NIMS-OS.

\begin{figure}
\centering
\includegraphics[scale=0.7]{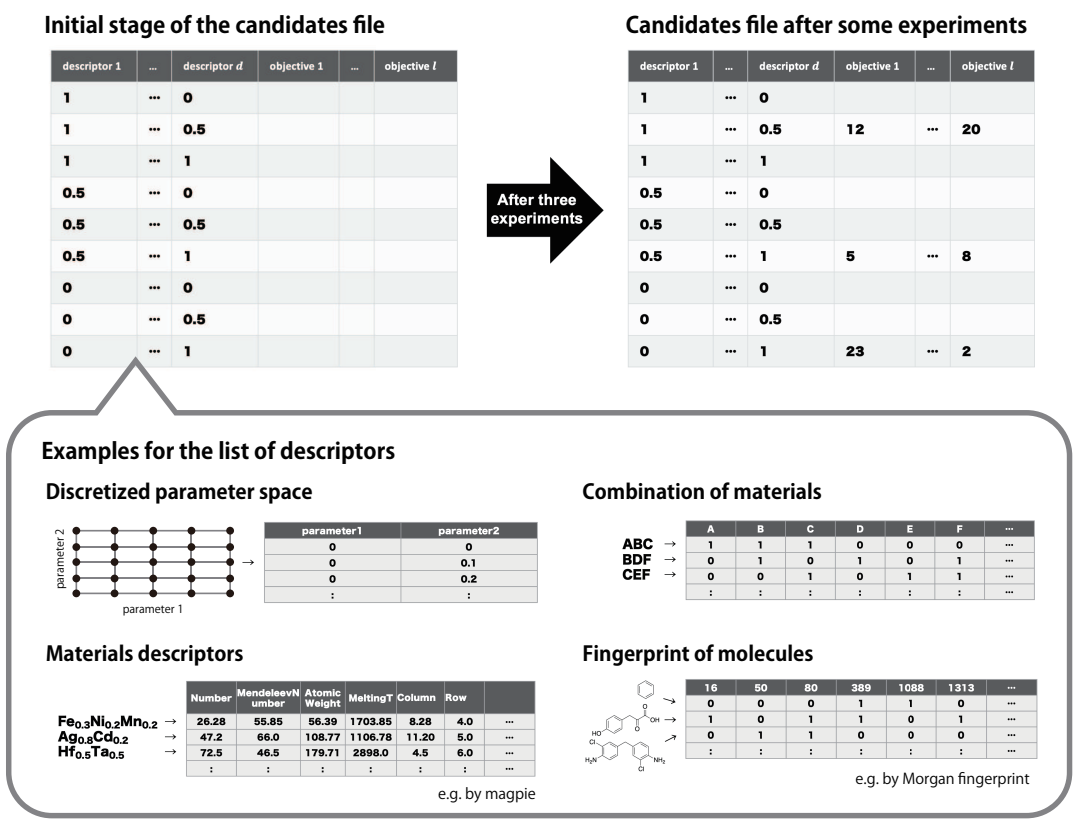}
\caption{
(Top panels) Examples of the candidates files of the initial stage and that after some experiments. 
Here, an example for the case that $N=9$ is shown.
(Bottom panels) Examples for the list of descriptors depending on the types of search space.
If the continuous parameter space is considered,
$D = \{ \mathbf{x}_i \}_{i=1,...,N}$ is the discretized parameters.
When the combination of materials is the search space,
the bit strings where the material used is represented by 1 and the material not used is represented by 0 is $D = \{ \mathbf{x}_i \}_{i=1,...,N}$.
Furthermore, materials descriptors from compositions obtained by such as magpie\cite{ong_python_2013,ward_general-purpose_2016} and fingerprint of molecules obtained by such as RDKit\cite{noauthor_rdkit_nodate} would be used as $D = \{ \mathbf{x}_i \}_{i=1,...,N}$.
}
\label{fig:candidates}
\end{figure}

The experimental condition is expressed as a real-valued vector $\mathbf{x}_i \in \mathbb{R}^d$. 
This condition is prepared with information such as the compositions and structures of materials and the processes required to synthesize them.
If the number of candidates for the experimental conditions is $N$,
the dataset for candidates is defined as $D = \{ \mathbf{x}_i \}_{i=1,...,N}$.
The initial candidates file is created by this dataset $D$.
An example of a candidates file with $l$ objective functions is presented in Fig.~\ref{fig:candidates}.
All the candidates of $D$ are written in the first $d$ columns. 
In this part, there should be no empty spaces. 
The next $l$ columns are used for the objective function values.
In this part, at the initial stage,
all cells are empty because experiments have not been performed for all the experimental conditions.

In NIMS-OS, 
some promising conditions are selected from among those listed in the candidates file using machine learning models (available algorithms are described in Section~\ref{subsec:ai_algo}).
When the values of objective functions are obtained by performing experiments,
the objective functions in the candidates file are updated, accordingly.
That is, when the experiments are completed for $M$ experimental conditions,
only results for $M$ conditions are entered at the $l$ columns for the objective functions.
Thus, at the next step, the experimental conditions are selected from among $N-M$ candidates.

\section{Modules in NIMS-OS} \label{sec:modules}

In this section,
we introduce the modules included in NIMS-OS for AI algorithms and robotic systems.
In the present work,
we prepared four and two types of modules as AI algorithms and robotic systems, respectively.

\subsection{AI algorithms} \label{subsec:ai_algo}

To select promising experimental conditions,
three types of AI algorithms are implemented as standard in NIMS-OS.
In addition,
random exploration can be selected.
Each algorithm is briefly explained in this subsection.
In the future, more algorithms will be made available.

\subsubsection{Bayesian optimization: PHYSBO}

Bayesian optimization (BO) is an optimization technique using machine learning prediction.
In this method, 
by using Gaussian process regression, the value of an objective function is predicted when the experimental conditions are input.
The next promising experimental conditions are then selected based on the prediction values.
Here, 
because the Gaussian process can evaluate not only the mean value of the prediction but also its variance,
an acquisition function defined by mean and variance can be used to make the selection.
In NIMS-OS, BO can be performed using the Python package PHYSBO\cite{MOTOYAMA2022108405}.
PHYSBO supports single- and multi-objective optimizations, and multiple proposals are calculated.
Note that the number of objective functions is recommended to be no more than three due to excessively large computational time with higher values.
In NIMS-OS, Thompson sampling is used to define the acquisition function for rapid calculation.
The key point in using PHYSBO is that the exploration is performed to maximize the objective functions. 
Thus, if a material with the smaller properties is explored, 
we need to add a negative value to the objective functions.

\subsubsection{Boundless objective-free exploration: BLOX}

BLOX is a Python package that performs boundless objective-free exploration.
It is based on an algorithm designed to select the next experimental conditions, to perform uniform sampling in the space of the objective functions.
For materials science, 
curious materials can be found using BLOX.
Specifically, BLOX trains machine learning models to predict objective functions from experimental conditions.
Experimental conditions that realize uniform sampling in the space of objective functions are found based on the Stein discrepancy evaluated using the prediction results.
In NIMS-OS, a modified version of the BLOX algorithm that can propose multiple candidates is implemented.
To select multiple candidates,
after the experimental condition with the largest Stein discrepancy is selected, 
another condition is selected when the predicted values of the selected condition are regarded as a correct value.
This procedure is iterated,
and we obtain multiple proposals.
In NIMS-OS, random forest regression is used as a prediction model.
Although BLOX can handle any number of objective functions, 
it is recommended that the number of the objective functions be limited to three or four, 
because exploration in more dimensions requires more time.
BLOX has been used to search chemical spaces\cite{D0SC00982B} and to explore superhard materials\cite{Ojih:2022vm}.

\subsubsection{Phase diagram construction: PDC}

PDC is a Python package that can create a detailed phase diagram with a small number of experiments.
To investigate a phase diagram, 
PDC proposes promising experimental conditions for the next experiment by using active learning.
Specifically, uncertainty sampling based on the label propagation method finds uncertain points in the phase diagram,
and these uncertain points are proposed for the next experiments.
PDC was developed to propose multiple experimental conditions for batch experiments\cite{doi:10.1080/27660400.2022.2076548}.
In NIMS-OS,
the least confident score is used as an uncertainty score to evaluate uncertain points.
Note that, for PDC, the objective function is the phase name or an index of phases,
and thus only a one-dimensional objective function can be specified in the candidates file.
PDC has been used to create new phase diagrams for the growth conditions of thin film\cite{doi:10.1021/acsmaterialslett.0c00104} and to determine large and small areas of creep phenomena in polymer materials\cite{doi:10.1080/14686996.2021.2025426}.

\subsubsection{Random exploration: RE}

In RE, the next candidate experimental condition is selected randomly.
This approach can be used to generate initial data before executing AI algorithms when no experimental data have yet been recorded.
Furthermore,
it can also be used to generate data for comparison as new AI algorithms can be developed.

\subsection{Robotic experiments}

The module for robotic experiments comprises two Python scripts. 
The first script creates input files for robotic experiments according to the experimental conditions selected by the AI and commands a robot to begin the experiment.
The second script analyzes the experimental results when the experiments are finished, 
and updates the candidates file.
At present,
two types of modules are implemented in NIMS-OS.

\subsubsection{Reference module for robotic experiments: STAN} \label{subsubsec:reference}

This module is a virtual implementation of the procedure for conducting robotic experiments. 
Thus, NIMS-OS can be run virtually, even without a robotic device.
In this module, the following steps are executed:
\begin{enumerate}
\item Create the input files for the robotic experiments in an appropriate folder according to the experimental conditions selected by the AI. 
In this reference module, we simply create a text file with a date as its name.

\item Send a signal to the robotic system to begin the experiments. 
Depending on the machine, various cases can be considered, such as sending a start signal via serial communication.
In this reference module, we assume that the experiments are begun by storing the inputend.txt file in the specified folder.

\item Wait until the robotic experiments are completed. 
This step includes various operations, such as receiving signals from the robot when the experiment is finished. 
This reference module assumes that the robot outputs outputend.txt file to indicate that the experiment is finished, and NIMS-OS continues waiting until this file appears.

\item Read the files of experimental results and extract the values of objective functions. 
Here, the case of simply reading results.csv, which contains the objective function values, is implemented.

\item Update the candidates file according to the values extracted in iv.
\end{enumerate}
Steps i and ii are performed by \texttt{preparation\_input.py}, and \texttt{analysis\_output.py} conducts steps iii-v.
In practice, for use with actual robotic systems, new modules can be created according to this reference module.

\subsubsection{NIMS automated robotic electrochemical experiments (NAREE) system: NAREE}

As a robotic system for materials science,
the NIMS Automated Robotic Electrochemical Experiments (NAREE) system\cite{Matsuda:2019aa,MATSUDA2022100832} can be used in NIMS-OS.
NAREE comprises a liquid-handling dispenser, an electrochemical measurement unit, 
and a robotic arm. 
By using a microplate-based electrochemical cell equipped with electrodes, 
the performance of electrolytes prepared by mixing solution by a liquid handling dispenser is electrochemically evaluated in a high-throughput manner.
This module was developed according to the procedures of reference module explained in Section~\ref{subsubsec:reference}.

\section{Usage of the NIMS-OS Python version} \label{sec:python}

\subsection{Install} \label{subsec:install}
NIMS-OS is written in Python3 programming language (version 3.6 or higher is required),
and it can be installed via PyPI as follows:
\begin{verbatim}
$ python3 -m pip install nimsos
\end{verbatim}
If this installation is successful, the following packages are also installed or updated automatically:
\begin{itemize}
\item \texttt{Cython}
\item \texttt{matplotlib}
\item \texttt{numpy}
\item \texttt{physbo}
\item \texttt{scikit-learn}
\item \texttt{scipy}
\end{itemize}

\subsection{Basic usage} \label{subsec:usage}

We show a small example program (Program~\ref{code:sample}) in which PHYSBO is performed.
In this program,
assuming no experimental results in the candidates file,
random exploration is performed in the first cycle.

\begin{lstlisting}[language=python, label=code:sample, caption=Small example of NIMS-OS for Bayesian optimization, frame=single, numbers=left, basicstyle=\footnotesize\ttfamily]
import nimsos

ObjectivesNum = 2
ProposalsNum = 2
CyclesNum = 3

candidates_file = "./candidates.csv"
proposals_file = "./proposals.csv"

input_folder = "./EXPInput"
output_folder = "./EXPOutput"

for K in range(CyclesNum):

    if K==0:
        method = "RE"
    else:
        method = "PHYSBO"

    nimsos.selection(method = method, 
                     input_file = candidates_file, 
                     output_file = proposals_file, 
                     num_objectives = ObjectivesNum, 
                     num_proposals = ProposalsNum)

    nimsos.preparation_input(machine = "STAN", 
                             input_file = proposals_file, 
                             input_folder = input_folder)

    nimsos.analysis_output(machine = "STAN", 
                           input_file = proposals_file, 
                           output_file = candidates_file, 
                           num_objectives = ObjectivesNum, 
                           output_folder = output_folder)
\end{lstlisting}

\subsubsection{Assignment of parameters and candidates file}

First, the parameters for closed-loop experiments are defined.
For example, when the number of objective functions is two, the number of proposals for each cycle is two, and the number of cycles is three. 
We define this in the code as follows:
\begin{verbatim}
ObjectivesNum = 2
ProposalsNum = 2
CyclesNum = 3
\end{verbatim}
Next, we specify a csv file containing the candidates of experimental conditions, which is prepared according to Section~\ref{sec:preparation}.
\begin{verbatim}
candidates_file = "./candidates.csv"
\end{verbatim}
The name of the file that will contain the experimental conditions selected by the AI is as follows:
\begin{verbatim}
proposals_file = "./proposals.csv"
\end{verbatim}

We specify the folder name where the input files for the robotic experiments are stored and the folder name where the results from the experiments are output, respectively, as follows:
\begin{verbatim}
input_folder = "./EXPInput"
output_folder = "./EXPOutput"
\end{verbatim}

\subsubsection{Execution of AI}

\texttt{nimsos.selection} is a class to select the next experimental conditions with the help of the AI.
For example, \texttt{nimsos.selection} is used as follows:
\begin{verbatim}
nimsos.selection(method = "PHYSBO", 
                 input_file = candidates_file, 
                 output_file = proposals_file, 
                 num_objectives = ObjectivesNum, 
                 num_proposals = ProposalsNum)
\end{verbatim}
The parameters of the \texttt{method} in this class indicate the module for AI algorithms.
For \texttt{method}, \texttt{"PHYSBO"} (Bayesian optimization), \texttt{"BLOX"} (objective free search), \texttt{"PDC"} (phase diagram construction), and  \texttt{"RE"} (random exploration) are specified.
The experimental conditions are selected from the data without the values of objective functions among \texttt{input\_file}.
In addition, 
selected conditions are outputted to \texttt{output\_file}.
For \texttt{num\_objectives}, the number of objectives is input,
and the number of proposals is specified as \texttt{num\_proposals}.
In general,
although many hyperparameters should be considered to use the AI,
they are determined automatically in NIMS-OS.
Note that if there are no experimental results in the candidates file,
only \texttt{"RE"} is used.
For \texttt{"PHYSBO"}, \texttt{"BLOX"}, and \texttt{"PDC"},
some values of objective functions must be stored in the candidates file.

\subsubsection{Preparation of input files for robotic experiments and execution of experiments}
\label{subsubsec:pre_input}

\texttt{nimsos.preparation\_input} is a class to prepare the input files for robotic experiments and send the start message to robot.
For example, \texttt{nimsos.preparation\_input} is used as follows.
\begin{verbatim}
nimsos.preparation_input(machine = "STAN", 
                         input_file = proposals_file, 
                         input_folder = input_folder)
\end{verbatim}
The parameter of \texttt{machine} selects the module of robotic experiments.
For \texttt{machine}, \texttt{"STAN"} which is the reference module for this procedure, and \texttt{"NAREE"} (NIMS automated robotic electrochemical experiments) are used.
For \texttt{input\_file}, the experimental conditions selected by the AI are specified.
In addition, the folder in the computer where the input files for robotic experiments are stored is referred to as the \texttt{input\_folder}.
In the \texttt{nimsos.preparation\_input} module,
the two functions \texttt{make\_machine\_file()}  and \texttt{send\_message\_machine()} should be modified depending on the robotic systems used.
The former creates the input files for robotic experiments from selected experimental conditions,
whereas later sends the message to begin the robotic experiments.

\subsubsection{Analysis of output files from experiments and update of candidates file}
\label{subsubsec:ana_output}

\texttt{nimsos.analysis\_output} is a class used to analyze the experimental results and update the candidates file.
For example, \texttt{nimsos.analysis\_output} is used as follows.
\begin{verbatim}
nimsos.analysis_output(machine = "STAN", 
                       input_file = proposals_file, 
                       output_file = candidates_file, 
                       num_objectives = ObjectivesNum, 
                       output_folder = output_folder)
\end{verbatim}
The parameter of \texttt{machine} is the same in \texttt{nimsos.preparation\_input} module,
which selects the module for robotic experiments.
Here, \texttt{"STAN"} and \texttt{"NAREE"} can be selected.
For \texttt{input\_file}, the experimental conditions selected by the AI are specified,
and \texttt{output\_file} is the name of the candidates file.
The file specified by \texttt{output\_file} is updated by this module.
In addition, for \texttt{num\_objectives}, the number of objectives is input.
For \texttt{output\_folder},
the folder in the computer where the results from robotic experiments are output is specified.
In the \texttt{nimsos.analysis\_output} module,
two functions \texttt{extract\_objectives()}  and \texttt{recieve\_exit\_message()} should be modified depending on the robot systems.
The former extracts the values of objective functions from the output files of robotic experiments,
and the later receives the message when the robotic experiments are finished.
If \texttt{"NAREE"} is selected,
\texttt{objectives\_info} should be specified as a dictionary indicating which objective function is extracted from the experimental results.

\subsection{Visualization of the results}

By using \texttt{nimsos.visualization},
the figures of the results are obtained.
When this module is used,
the new folder named \texttt{fig} is prepared in advance in the same folder where the main script is stored.
The figures are output to this folder.
\texttt{nimsos.visualization.plot\_history} and \texttt{nimsos.visualization.plot\_distribution.plot} create figures for the history and distributions of objective functions, respectively.
These modules are useful when using AI algorithms other than PDC. 
In contrast,
\texttt{nimsos.visualization.plot\_phase\_diagram.plot} creates the predicted phase diagram when PDC is used as an AI algorithm.

\section{Usage of the NIMS-OS GUI version} \label{sec:gui}

\begin{figure}
\centering
\includegraphics[scale=1]{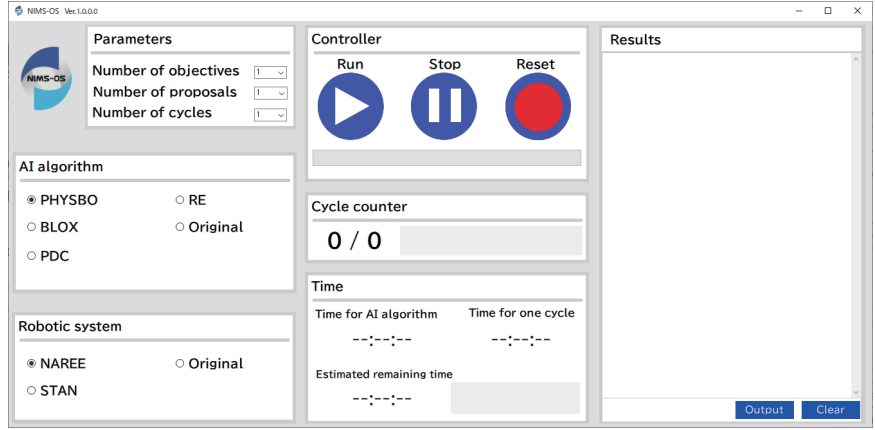}
 \caption{
Operation screen of the NIMS-OS GUI version.
}
\label{fig:gui}
\end{figure}

A GUI version of NIMS-OS has been developed for easy execution,
which is available at \texttt{https://github.com/nimsos-dev/nimsos-gui}.
This can be used after installing the required Python version, as described in Section~\ref{subsec:install}, 
and performing the installation is described in the manual (\texttt{https://nimsos-dev.github.io/nimsos/docs/en/index.html}).
Figure~\ref{fig:gui} shows the operation screen of the NIMS-OS GUI version.
In this GUI version, 
the name of the candidates file is fixed to candidates.csv, 
and the name of the proposals file is fixed to proposals.csv.
The execution procedure is as follows:
\begin{enumerate}

\item Specify the number of objectives, proposals, and cycles in the \textit{Parameters} section.

\item Select the method to be used in the \textit{AI algorithm} section. 
If we use a newly created module for AI method named \texttt{ai\_tool\_original.py}, 
click on Original.

\item Select the robotic system in the \textit{Robotic system} section.
If we use a newly created module for robotic systems named \texttt{preparation\_input\_original.py} and \texttt{analysis\_output\_original.py}, 
click on Original.

\item Press the ``run'' button on the \textit{Controller} section to begin the automated materials exploration.
\end{enumerate}
When NIMS-OS is started, the \textit{Cycle counter} begins to operate.
Furthermore, in the \textit{Time} section, 
the amount of time required to execute the AI algorithm and a single cycle are measured,
and the remaining time is also output.
The standard output of the Python version is displayed in real time in the \textit{Results} section, 
and these output results can be saved as a file by pressing the \textit{Output} button.
In addition, to pause the automated exploration, 
the user can press the stop button of the \textit{Controller} section.
Note that pressing this button does not stop the process immediately, 
but when the candidates file is updated, NIMS-OS is stopped.
To reset the settings, press the reset button on the \textit{Controller}.
The operation with NAREE is shown as a video of Supplemental Movie 1.

\section{Application} \label{sec:app}

To demonstrate the effectiveness of NIMS-OS for the application of automated robotic experiments, we applied NIMS-OS for NAREE system and performed an exploration for multi-component electrolytes that maximize the performance of lithium metal electrode.
The anode-free type microplate based electrochemical cells were fabricated using LiFePO$_4$ as positive electrode and Cu foil as negative electrode. 
The cells were subjected to charging process with capacity limitation of 0.05 mAh. 
After that, the cells were subjected to discharge process. 
Here, we defined the discharge time as one-dimensional objective function. 
In this case, the longer discharge time represents the better battery performance (higher capacity). 
Using such experimental setup, a combination of electrolyte additives was optimized to maximize the discharge time. 
Five different additives were selected from a list of 16 compounds (Table~\ref{tab: additives}) and injected into electrochemical cell containing 1 M LiTFSI in TEGDME.
In this case, the number of candidates for combination of electrolyte additives is $_{16}{\rm C}_5 = 4,368$.
The candidate files for this experiment were prepared in similar manner as shown in Fig.~\ref{fig:candidates} (combination of materials). 
In our experiment, 32 electrochemical cells were prepared in one microplate and 32 experiments were parallelly performed in 2 hours.

\begin{figure}
\centering
\includegraphics[scale=0.3]{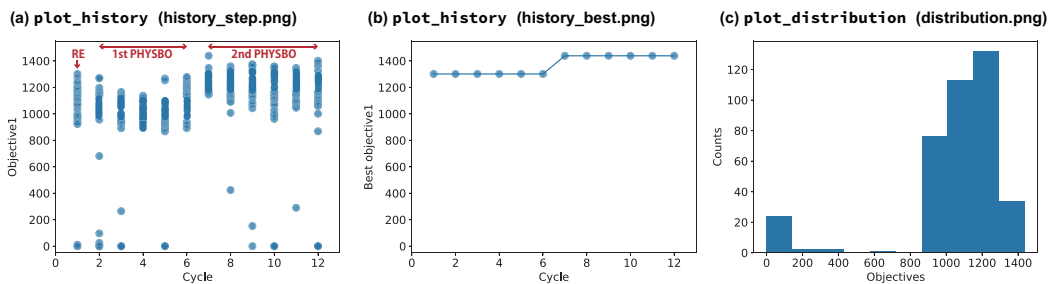}
\caption{
Output results from NIMS-OS for automated exploration for electrolytes using NAREE system:
(a) history\_step.png and (b) history\_best.png by \texttt{nimsos.visualization.plot\_history} and (c) distribution.png by \texttt{nimsos.visualization.plot\_distribution.plot}.
The target property is the discharge time and its unit is sec.
In the first cycle, RE is used to generate initial states.
After the second cycle, PHYSBO is used. 
}
\label{fig:exp}
\end{figure}

For the autonomous experiments for searching multi-component electrolytes using NAREE system operated by NIMS-OS, at first, 32 parallel experiments (one microplate) were performed by random exploration using \texttt{"RE"} because we do not have initial data at this stage. 
After obtaining the initial data by \texttt{"RE"}, next five cycles of experiments (five microplates) were performed by BO using \texttt{"PHYSBO"}. 
Notably, fully automated experiment was continuously conducted without any human intervention for 10 hours. 
After that, addition six cycles of experiments (six microplates) were also performed by BO. 
In total, 384 experiments were performed.
The obtained results can be visualized by using \texttt{nimsos.visualization} in the Python version of NIMS-OS, as already mentioned in Section 4.3, and the time course of objective function and the histogram distribution of the results in the total 384 experiments were summarized in Fig.~\ref{fig:exp}. 
The results clearly revealed that the best electrolyte composition was discovered at 7th experimental cycle. 
In Table~\ref{tab: rank}, the details of electrolyte composition for top 10 samples that enhanced the discharge time were summarized. 
The electrolyte containing, 100 mM LiPF$_6$, 100 mM LiTFSI, 2 vol.\% PC, 2 vol.\% FEC, and 2 vol.\% VC, exhibits the highest discharge time of 1439.09 sec. 
It should be noted that the possible maximum discharge time is 1800 sec since the current density during discharge was set to 0.1 mA. 
Thus, there is still much room for improvement of battery performance. 
In addition, there can be seen that most of the top 10 samples contains VC and/or FEC. 
These results are essentially consistent with the knowledge in this filed that VC and FEC has positive effect for improving the performance of the lithium metal electrode \cite{OTA2004565,https://doi.org/10.1002/adfm.201605989}.

\begin{table}
\centering
\caption{
List of 16 types of additives used in an automated exploration for new electrolytes using NAREE system.
For all additives, the solvent is fixed as TEGDME.
}
\label{tab: additives}
\begin{tabular}{|l|ll|}
\hline
ID & Additive & Concentration \\
\hline
1 & lithium bis(pentafluoroethanesulfonyl)imide (LiBETI) & 100 mM \\
2 & LiPF$_6$ & 100 mM \\
3 & LiBF$_4$ & 100 mM \\
4 & lithium bis(trifluoro methanesulfonyl)imide (LiTFSI) & 100 mM \\
5 & LiTfO & 100 mM \\
6 & LiClO$_4$ & 100 mM \\
7 & lithium bis(oxalate)borate (LiBOB) & 10 mM \\
8 & LiAsF$_6$ & 10 mM \\
9 & LiF & 10 mM \\
10 & N-methyl-2-pyrrodione (NMP) & 2 vol.\% \\
11 & sulfolane & 2 vol.\% \\
12 & dimethyl sulfoxide (DMSO) & 2 vol.\% \\
13 & propylene carbonate (PC) & 2 vol.\% \\
14 & ethylene carbonate (EC) & 2 vol.\% \\
15 & fluoroethylene carbonate (FEC) & 2 vol.\% \\
16 & vinylene carbonate (VC) & 2 vol.\% \\
\hline
\end{tabular}
\end{table}

\begin{table}
\centering
\caption{
Top 10 compositions that enhanced the discharge time.
The found cycle number is also shown.
}
\label{tab: rank}
\resizebox{\columnwidth}{!}
{\begin{tabular}{|l|lllll|l|l|}
\hline
Ranking & Additive 1 & Additive 2 & Additive 3 & Additive 4 & Additive 5  & Discharge time & Found cycle \\
\hline
1 & 100 mM LiPF$_6$ & 100 mM LiTFSI & 2 vol.\% PC & 2 vol.\% FEC & 2 vol.\% VC & 1439.09 sec. & 7th \\
2 & 100 mM LiBETI & 100 mM LiTfO & 10 mM LiBOB & 2 vol.\% EC & 2 vol.\% FEC & 1401.97 sec. & 12th \\
3 & 2 vol.\% NMP & 2 vol.\% sulfolane & 2 vol.\% DMSO & 2 vol.\% PC & 2 vol.\% FEC & 1374.86 sec. & 9th \\
4 & 100 mM LiBF$_4$ & 100 mM LiTFSI & 100 mM LiTfO & 10 mM LiF & 2 vol.\% FEC & 1365.57 sec. & 9th \\
5 & 100 mM LiBETI & 100 mM LiBF$_4$ & 10 mM LiBOB & 2 vol.\% PC & 2 vol.\% FEC & 1364.32 sec. & 12th \\
6 & 100 mM LiTFSI & 100 mM LiTfO & 10 mM LiAsF$_6$ & 2 vol.\% FEC & 2 vol.\% VC & 1358.99 sec. & 10th \\
7 & 100 mM LiBETI & 10 mM LiBOB & 10 mM LiF & 2 vol.\% EC & 2 vol.\% VC & 1357.43 sec. & 10th \\
8 & 100 mM LiBETI & 100 mM LiTFSI & 10 mM LiBOB & 10 mM LiF & 2 vol.\% FEC & 1356.39 sec. & 9th \\
9 & 100 mM LiBF$_4$ & 100 mM LiTFSI & 100 mM LiClO$_4$ & 2 vol.\% sulfolane & 2 vol.\% VC & 1347.23 sec. & 11th \\
10 & 100 mM LiPF$_6$ & 100 mM LiTfO & 10 mM LiAsF$_6$ & 10 mM LiF & 2 vol.\% EC & 1346.72 sec. & 7th \\
\hline
\end{tabular}
}
\end{table}

\section{Summary} \label{sec:summary}

In this study,
we developed NIMS-OS to implement a closed loop of robotic experiments and AI for automated materials exploration. 
We anticipate that this software can serve as a generic control system.
To use NIMS-OS, a candidates file listing experimental conditions as a materials search space should be prepared in advance.
This allows various problems for automated materials exploration to be commonly performed in NIMS-OS.
Establishing standards for automated materials exploration is a key advantages of such generic control software.
Using NIMS-OS and the NAREE system, 
we have also demonstrated an example of automatic exploration for electrolytes.

At present, 
NIMS-OS does not include a sufficient set of available AI algorithms and robotic experimental systems.
For the further growth of this OS, developing and releasing more modules for various AI algorithms and robotic systems will be essential.
In automated materials exploration, more experimental data are generated compared to human experiments.
Thus,
the ability to store and utilize experimental data should be implemented as an extension in NIMS-OS
We will continue to enhance the extensions available for NIMS-OS to develop it as a game changer for digital transformation (DX) in materials science.


\section*{Acknowledgements}
The authors thank Masahiko Demura, Hideki Yoshikawa, and Masanobu Naito for valuable discussions.
The authors also thank Kazuha Nakamura for experimental contributions,
and thank Satoshi Murata, Daisuke Ryuno, and Hiromichi Taketa
for the development of NIMS-OS.
The research presented in this article was supported by the MEXT Program: Data Creation and Utilization-Type Material Research and Development Project (Grant Numbers JPMXP1122712807).


\bibliographystyle{tfnlm}
\bibliography{nimbos-bib}

\end{document}